\documentclass[prb,twocolumn,showpacs,superscriptaddress]{revtex4}
\usepackage{amssymb}
\usepackage{amsmath}
\usepackage{graphicx}

\usepackage{rotating}
\usepackage{graphicx}
\begin{document}


\title{Anomalous flux avalanche morphology in a $a$-MoGe superconducting film with
a square antidot lattice - experiment and simulation}

\author{M. Motta}
\author{F. Colauto}
\affiliation{Departamento de F\'{\i}sica, Universidade Federal de S\~{a}o Carlos, 13565-905 S\~{a}o Carlos, SP, Brazil}

\author{W. A. Ortiz}
\affiliation{Departamento de F\'{\i}sica, Universidade Federal de S\~{a}o Carlos, 13565-905 S\~{a}o Carlos, SP, Brazil}
\affiliation{Centre for Advanced Study, Norwegian Academy of Science and Letters, NO-0271, Oslo, Norway}

\author{J. I. Vestg{\aa}rden}
\affiliation{Department of Physics, University of Oslo, POB 1048, Blindern, 0316 Oslo, Norway}

\author{T. H. Johansen}
\affiliation{Centre for Advanced Study, Norwegian Academy of Science and Letters, NO-0271, Oslo, Norway}
\affiliation{Department of Physics, University of Oslo, POB 1048, Blindern, 0316 Oslo, Norway}
\affiliation{Institute for Superconducting and Electronic Materials, University of Wollongong, Northfields Avenue, Wollongong, NSW 2522, Australia}

\author{Jo Cuppens}
\author{V. V. Moshchalkov}
\author{A. V. Silhanek}

\affiliation{INPAC -- Institute for Nanoscale Physics and Chemistry, Nanoscale Superconductivity\\ and Magnetism Group, K.U.Leuven, Celestijnenlaan 200D,
B--3001 Leuven, Belgium}


\date{\today}
\begin{abstract}
 
We have employed magneto-optical imaging to visualize the occurrence of flux avalanches in a superconducting film of $a$-MoGe. The specimen was decorated with square antidots arranged in a square lattice. We observed avalanches with the anomalous habit of forming trees where the trunk is perpendicular to the main axis of the square lattice, whereas the branches form angles of 45 degrees. The overall features of the avalanches, and in particular the 45 degree direction of the branches, were confirmed by numerical simulations.

\end{abstract}


\maketitle


Flux avalanches triggered by thermomagnetic instabilities have been reported in a variety of superconducting films of Nb~\cite{Duran95},
MgB${_2}$~\cite{Johansen01,Johansen02}, Nb${_3}$Sn~\cite{Rudnev03}, NbN~\cite{Rudnev05}, YNi${_2}$B${_2}$C~\cite{{Wimbush04}}, and YBCO~\cite{Leiderer93}.
Such events, in which flux bursts suddenly invade the sample, have also been studied in superconducting films decorated with a lattice of antidots,
which guide flux motion, as revealed by magneto-optical imaging (MOI) experiments~\cite{Vlasko-Vlasov00,Menghini05}.

\begin{figure}
\begin{center}
\includegraphics[width=7cm]{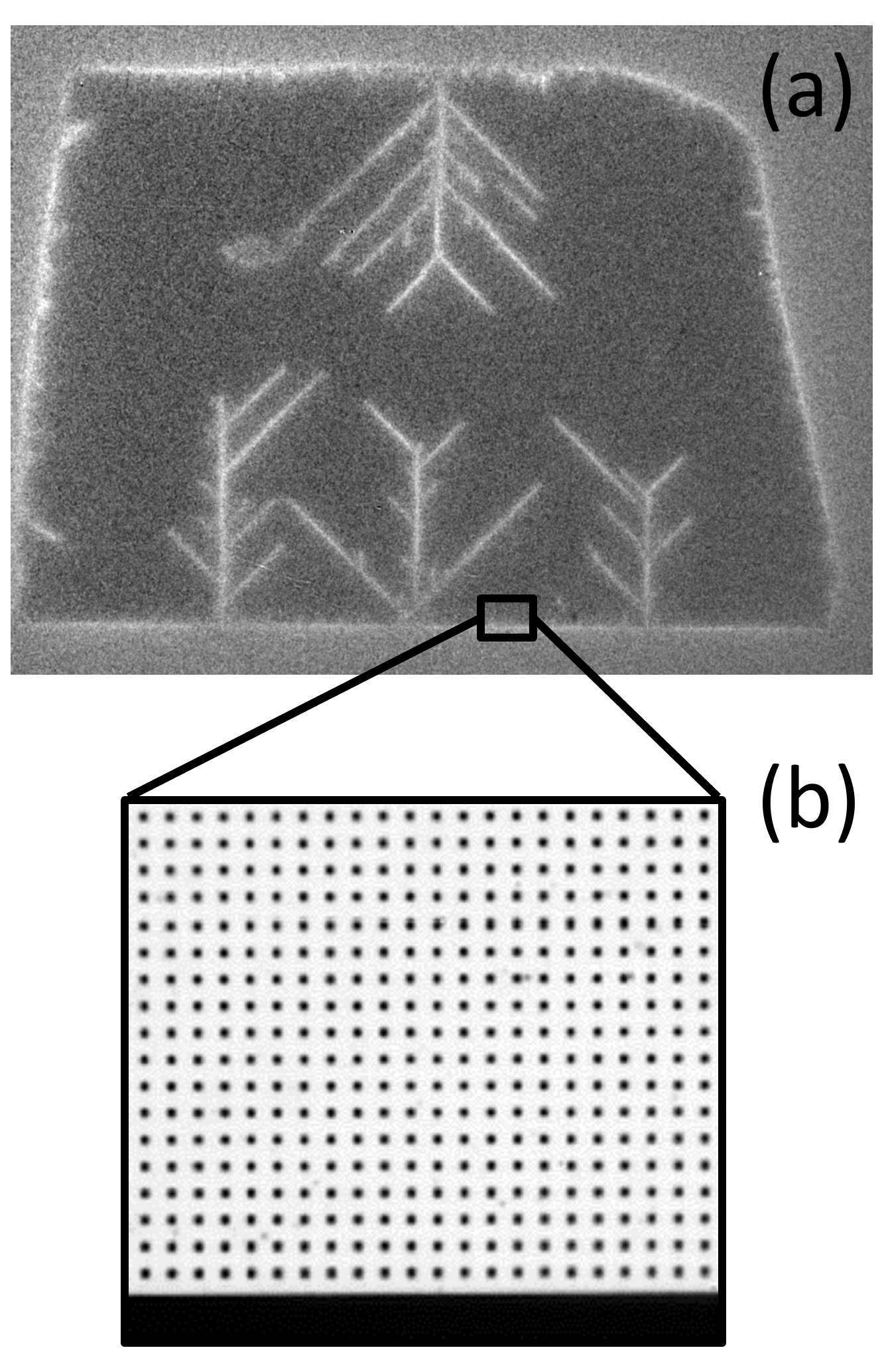}
\caption{(a) Magneto optical image of flux penetration at T = 4.5 K and H = 1 Oe, applied perpendicular to the film. The image brightness represents the magnitude of the local flux density. (b) Optical image of the square lattice of square antidots.}
\end{center}
\end{figure}


The sample investigated in the present work is an amourphous film of MoGe ($a$-MoGe) approximately rectangular in shape, with thickness $t=25$ nm and lateral dimensions roughly $2.0$ mm x $2.6$ mm. A square lattice of square antidots (ADs) was fabricated by electron beam lithography. The $a$-MoGe was deposited via pulsed laser deposition from a target of Mo$_{78}$Ge$_{22}$ with purity 99.96$\%$, on top of a SiO$_2$ insulating substrate. The patterned sample has square antidots with sides 0.4 $\mu$m. The period of the pattern is $w=1.5~\mu$m, which corresponds to a commensurability field 
$H_1=\frac{\Phi_0}{w^2} =$ 9.2 Oe at which the densities of vortices and antidots match each other~\cite{commensurability comment}. Using the expression
for the upper critical field and for the dirty limit~\cite{Schmidt97}, the zero temperature superconducting coherence length, $\xi_{GL}(0)=5$ nm was determined, and the
penetration depth, $\lambda_{GL}(0)=517$ nm was obtained for a similar plain film. Figure 1b shows the orientation of the antidot lattice relative to the edges of the sample.

\begin{figure}
\begin{center}
\includegraphics[width=7cm]{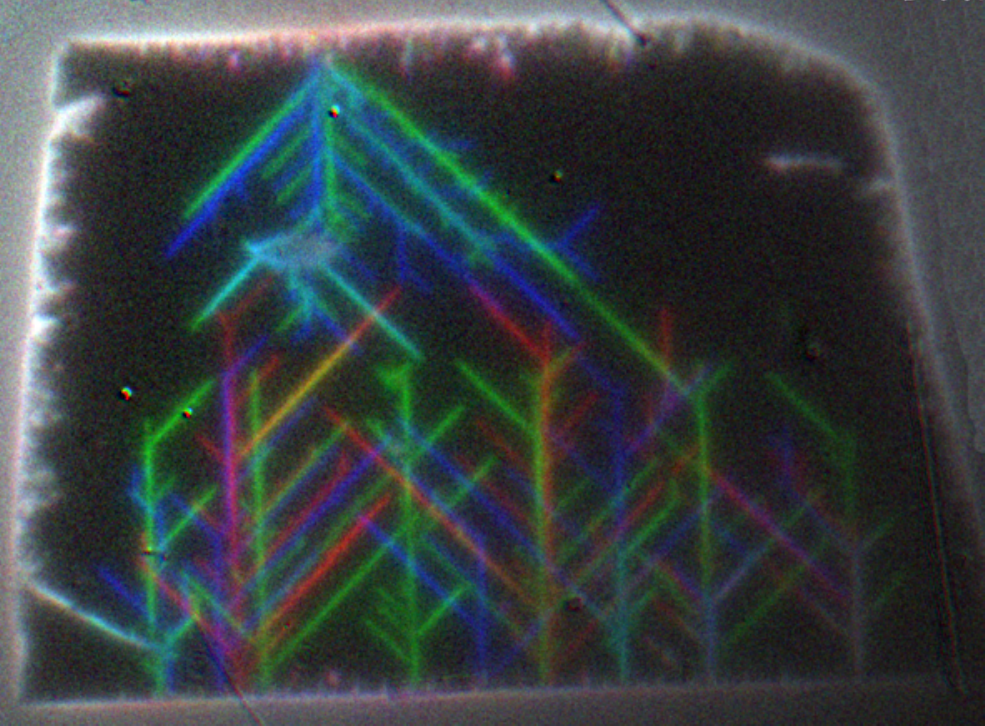}
\caption{Overlay of 3 repeated identical MOI experiments showing the flux penetration into the $a$-MoGe film, carried out at T = 3 K with H = 1.4 Oe. Each experiment is color coded in red, green and blue, thus the presence of these colors in the overlay shows that the penetration is non-reproducible. }
\end{center}
\end{figure}

Characterization measurements of the ac susceptibility and dc magnetization were carried out in commercial Quantum Design
equipments (MPMS and PPMS) showing that the superconducting transition temperature at zero field is $T_c=6.74$ K.
 The MOI technique employed relies on the Faraday effect ~\cite{Helseth01} of a
garnet indicator film placed on top of the superconducting specimen. The indicator used in the present work was a Bi-substituted yttrium iron garnet film
(Bi:YIG) with in-plane magnetization.

Figure 1a shows a magneto optical image taken at T = 4.5 K and an applied DC field of H = 1 Oe. Noticeably, the main trunk of the tree-like avalanches are predominantly perpendicular to one of the sample edges whereas, surprisingly, the branches form angles of 45 degrees relative to the main axes of the antidot lattice. Avalanches directed at 45 degrees have previously been observed only in a Nb film and there only from edges strongly inclined relative to the main axis of the antidot lattice~\cite{Vlasko-Vlasov00}. 

Figure 2 shows the results of three different experiments, all of them carried out at T = 3 K with H = 1.4 Oe. The images are colored red, green and blue, so that overlapping penetrations in all three experiments appear as shades of gray, whereas non-repetitive flux patterns combine in color. Evidently, there is essentially no overlap in the dendritic flux patterns formed in the three experiments. The background flux penetration, however, is fully reproducible. 

\begin{figure}[bbb]
\begin{center}
\includegraphics[width=7cm]{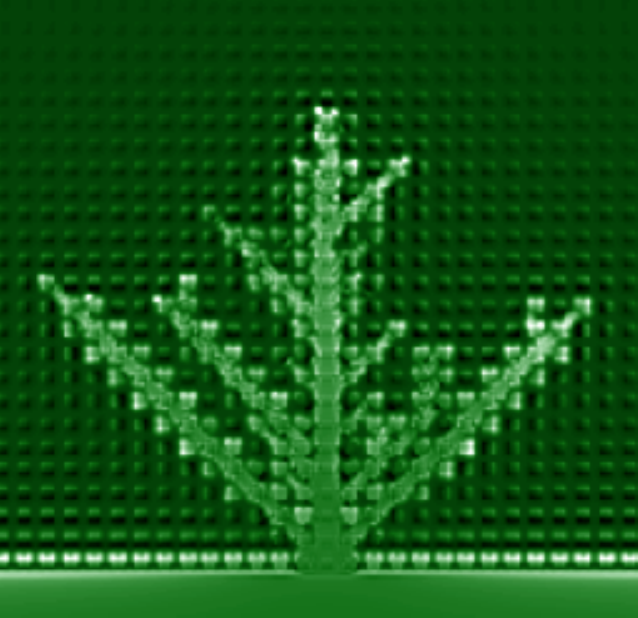}
\caption{Results of numerical calculations of the flux penetration into a superconducting film with a square array of square antidots where a thermomagnetic feedback is taken into account.
Details of the numerial procedure will be published elsewhere. }
\end{center}
\end{figure}

The origin of dendritic avalanches in superconducting films is a thermomagnetic instability mechanism
between the Joule heating created by vortex motion and the reduction
of the critical current density as temperature
increases. The instability is also a consequence of the 
nonlinear material characteristics of type II superconductors, 
which is conventionally approximated by a power law
\begin{equation}
  \label{EJ}
  \mathbf E = \frac{\rho_0}{d}\left(\frac{J}{J_c}\right)^{n-1}\mathbf J
  ,
\end{equation}
where $\mathbf E$ is electric field, $\mathbf J$ is sheet current, $J=|\mathbf J|$,
$\rho_0$ is a resistivity constant,  $d$ is sample thickness, 
$J_c$ is critical sheet current, and $n$
is the creep exponent. The temperature dependencies are taken as 
\begin{equation}
  J_c=J_{c0}(1-T/T_c),~~~ n=n_1/T,
\end{equation}
where $T_c$ is the critical temperature. The electrodynamics must be supplemented 
by the heat diffusion equation 
\begin{equation}
  \label{dotT}
  c\dot T = \kappa \nabla^2T-\frac{h}{d}(T-T_0)+\frac{1}{d}JE
  ,
\end{equation}
where $c$ is specific heat, $\kappa$ is thermal conductivity, $h$
is the coefficient for heat removal to the substrate, 
and $T_0$ is the substrate temperature. 

Eq.~\eqref{dotT} must be solved together with Maxwell's equations and
the material law, Eq.~\eqref{EJ}.  The description of the simulation
method is found in Ref.~\onlinecite{vestgarden11}, so here only the key
ingredients are outlined.  The main challenge is to invert the
Biot-Savart law in an efficient way. This is done by including also
the vacuum outside the sample in the simulation formalism. At the cost
of including the extra space, one can use a real space/Fourier space
hybrid method with the very attractive performance scaling of $O(N\log
N)$, where $N$ is the number of grid points.

The area of the sample was discretized on a 512$\times$512 grid, 
with the geometry of strip, emulated by periodic 
boundary conductions. A total number of 43$\times$43 square antidots 
where imposed,  so that the antidot coverage was 1/4 of the total area. 
The initial state 
of the simulation was prepeared by ramping the external
field with thermal feedback turned off, so that a critical state 
was formed from the edges. When later, the thermal feedback was turned 
on, and at the same time a local spot close to the edge was heated
above $T_c$, a flux avalanche developed from 
the heated spot.

Figure 3 shows a map of the local flux density obtained by the 
numerical simulation. The avalanche of the figure demonstrates 
a pronounced guidance effect, contrary to the case of plain films or 
films patterned with random disorder. Just as in the experiments, 
there is a main trunk perpendicular to the edge and two main branches
forming an angle of 45 degrees with both the edge and the directions 
of the antidot lattice.


In summary, we have employed MOI to observe flux avalanches in superconducting films of $a$-MoGe decorated with a square lattice of square antidots. Avalanches have the form of trees where the main trunk is perpendicular to the sample edge, whereas its branches form an angle of 45 degrees with the main axes of the antidot lattice. The overall features of the avalanches, and in particular the 45 degree direction of the branches, were confirmed by numerical simulations.


This work was partially supported by the Methusalem Funding of the Flemish Government, the NES-ESF program, the Belgian IAP, the Fund for Scientific
Research-Flanders (FWO-Vlaanderen), the UK Engineering and Physical Sciences Research Council and by the Brazilian funding agencies FAPESP and CNPq. AVS is
grateful for the support from the FWO-Vlaanderen. THJ acknowledges the financial support of the Norwegian Research Council.
\\


\begin{thebibliography}{99}

\bibitem{Duran95}
C. A. Duran et al., Phys. Rev. B {\bf 52}, 75 (1995).

\bibitem{Johansen01}
T.H. Johansen, M. Baziljevich, D.V. Shantsev, P.E. Goa, Y.M. Galperin, W.N. Kang, H.J. Kim, E.M. Choi, M-S. Kim, S.I. Lee, Supercond. Sci.  Tech. 14 (2001)
726-728.

\bibitem{Johansen02}
T.H. Johansen, M. Baziljevich, D.V. Shantsev, P.E. Goa1, Y.M. Galperin, W.N. Kang, H.J. Kim, E.M. Choi, M.-S. Kim, S. I. Lee, Europhys. Lett. {\bf 59}, 599
(2002).

\bibitem{Rudnev03}
I.A. Rudnev, S.V. Antonenko, D.V. Shantsev, T.H. Johansen, A.E. Primenko, Cryogenics {\bf 43}, 663 (2003).

\bibitem{Rudnev05}
I.A. Rudnev, D.V. Shantsev, T.H. Johansen, A.E. Primenko, Appl. Phys. Lett. {\bf 87}, 042502 (2005).

\bibitem{Wimbush04}
S. C. Wimbush, B. Holzapfel, and C. Jooss, J. Appl. Phys. {\bf 96}, 3589 (2004).

\bibitem{Leiderer93}
P. Leiderer et al., Phys. Rev. Lett. {\bf 71}, 2646 (1993).

\bibitem{Hebert03}
S. Hebert, L. Van Look, L. Weckhuysen, V. V. Moshchalkov, Phys. Rev. B 67 (2003) 224510.

\bibitem{Vlasko-Vlasov00}
V. Vlasko-Vlasov, U. Welp, V. Metlushko, and G. W. Crabtree, Physica C 341-348 (2000) 1281.

\bibitem{Menghini05}
M. Menghini, R. J. Wijngaarden, A. V. Silhanek, S. Raedts, and V. V. Moshchalkov, Phys. Rev. B 71, 104506 (2005).

\bibitem{commensurability comment}
Matching effects are less pronounced at the temperature interval reported here (i.e., not very close to $T_c$). The commensurability field, characteristic
of the lattice of ADs, is a reference for the window of magnetic fields at which the events treated here take place.

\bibitem{Schmidt97}
V.V. Schmidt, The Physics of Superconductors, edited by P. Muller and A. V. Ustinov (Springer, Berlin-Heidelberg (1997).

\bibitem{Helseth01}
L. E. Helseth et al., Phys. Rev. B 64, 174406 (2001)

\bibitem{vestgarden11}
J. I. Vestgarden et al., Phys. Rev. B 84, 054537 (2011)

\end{thebibliography}
\end{document}